\documentclass[conference,a4paper]{IEEEtran}
\ifCLASSINFOpdf
   \usepackage[pdftex]{graphicx}
   \graphicspath{{Images/}}
\else
\fi
\usepackage{amssymb}
\usepackage[cmex10]{amsmath}

\usepackage{mdwmath}
\usepackage{mdwtab}

\usepackage[tight,footnotesize]{subfigure}
\usepackage[nolist]{acronym}

\begin{document}

\begin{acronym}
\acro{A/A}{Air-to-air}%
\acro{A-EXEC}{Auto Execute COCR Service}%
\acro{AAC}{Airline Administrative Control}%
\acro{ACK}{Acknowledgement}%
\acro{ACP}{Aeronautical Communications Panel}%
\acro{AF}{Assured Forwarding}%
\acro{AL}{Application Layer}%
\acro{ANSP}{Air Navigation Service Provider}%
\acro{AOC}{Airline Operational Control}%
\acro{APC}{Air Passenger Communication}%
\acro{APT}{Airport}%
\acro{ATC}{Air Traffic Control}%
\acro{ATI}{Access Technology Independent}%
\acro{ATM}{Air Traffic Management}%
\acro{ATN}{Aeronautical Telecommunication Network}%
\acro{ATN/IPS}{Aeronautical Telecommunication Network / Internet Protocol Suite}%
\acro{ATN/OSI}{Aeronautical Telecommunication Network / Open Systems Interconnection Reference Model}%
\acro{ATS}{Air Traffic Services}%
\acro{ATSC}{Air Traffic Service Communication}%
\acro{AVBDC}{Absolute Volume Based Dynamic Capacity}%
\acro{AWGN}{Additive White Gaussian Noise}%
\acro{BB}{Bandwidth Broking} %
\acro{BBFRAME}{Baseband frame}%
\acro{BE}{Best Effort}%
\acro{BER}{Bit Error Rate}%
\acro{BGAN}{Broadband Global Area Network}%
\acro{BSM}{Broadband Satellite Multimedia}%
\acro{CA}{Constrained Architecture}%
\acro{CAC}{Connection Admission Control}%
\acro{CC}{Congestion Control}%
\acro{CDMA}{Code Divison Multiple Access}%
\acro{CLNP}{Connectionless Networking Protocol}%
\acro{CLTP}{Connectionless Transport Protocol}%
\acro{CoA}{Care of Address}%
\acro{CoS}{Class of Service}%
\acro{COCR}{Communications Operating Concept and Requirements for the Future Radio System}%
\acro{COTP}{Connection Oriented Transport Protocol}%
\acro{COTS}{Commercial-Off-The-Shelf}%
\acro{CRA-CC}{CRA-Convolutional Code}%
\acro{CRA-SH}{CRA-Shannon Bound}%
\acro{CRA}{Contention Resolution ALOHA}%
\acro{CRC}{Cyclic Redundancy Check}%
\acro{CRDSA}{Contention Resolution Diversity Slotted ALOHA}%
\acro{CRDSA++}{Contention Resolution Diversity Slotted ALOHA++}%
\acro{CS}{Class Selector}%
\acro{DAMA}{Demand Assigned Multiple Access}%
\acro{DiffServ}{Differentiated Services}%
\acro{DoD}{Department of Defence}%
\acro{DSA}{Diversity Slotted ALOHA}%
\acro{DSCP}{DiffServ Code Point}%
\acro{DVB}{Digital Video Broadcasting}%
\acro{DVB-RCS+M}{Digital Video Broadcasting - Return Channel via Satellite Mobility}%
\acro{DVB-RCS}{Digital Video Broadcasting - Return Channel via Satellite}%
\acro{DVB-S2}{Digital Video Broadcasting - Second Generation}%
\acro{DVB-S2/RCS}{Digital Video Broadcasting - Second Generation / Return Channel via Satellite}%
\acro{ECN}{Explicite congestion notification}%
\acro{ECRA}{Enhanced Contention Resolution ALOHA}%
\acro{ECTL}{Eurocontrol}%
\acro{EF}{Expedited Forwarding}%
\acro{ENR}{En-Route}%
\acro{ES}{End System}%
\acro{ESA}{European Space Agency}%
\acro{ETSI}{European Telecommunications Standards Institute}%
\acro{ETSI-BSM}{European Telecommunications Standards Institute - Braodband Satellite Multimedia}%
\acro{FAA}{Federal Aviation Administration}%
\acro{FCA}{Free Capacity Assignment}%
\acro{FDM}{Frequency Division Multiplexing}%
\acro{FEC}{Forward Error Correction}%
\acro{FET}{First Entry Times}%
\acro{FL}{Forward Link}%
\acro{GEO}{Geostationary Orbit}%
\acro{GIG}{Global Information Grid}%
\acro{GSE}{Generic Stream Encapsulation}%
\acro{HA}{Home Agent}%
\acro{HF}{High Frequency}%
\acro{IC}{Interference Cancellation}%
\acro{ICAO}{International Civil Aviation Organization}%
\acro{IETF}{Internet Engineering Task Force}%
\acro{IntServ}{Integrated Services}%
\acro{IP}{Internet Protocol}%
\acro{IPS}{Internet Protocol Suite}%
\acro{IPSec}{IP Security}%
\acro{IRA}{Irregular Repetition contention resolution ALOHA}%
\acro{IRCRA}{Irregular Repetition Contention Resolution ALOHA}%
\acro{IRSA}{Irregular Repetition slotted ALOHA} %
\acro{IS}{Intermediate System}%
\acro{IS-IS-TE}{Intermediate System to Intermediate System extension for Traffic Engineering}%
\acro{ISI}{Input Stream Identifier}%
\acro{ISO/OSI}{International Organization for Standardization/Open Systems Interconnection}%
\acro{Ku}{Kurz-under}%
\acro{Ka}{Kurz-above}%
\acro{KL-d}{Kullback-Leibler divergence}
\acro{LAE}{Link Access Equipment}%
\acro{LAN}{Local Area Network}%
\acro{LD}{Low Density}%
\acro{LDACS}{L-band Digital Aeronautical Communication System}%
\acro{L-DACS}{L-band Digital Aeronautical Communication System}%
\acro{LDPC}{Low Density Parity Check Codes}%
\acro{LEO}{Low Earth Orbit}%
\acro{MAC}{Medium Access}%
\acro{MCAST}{Multicast}%
\acro{MF-TDMA}{Multifrequency Time Division Multiple Access}%
\acro{MH}{Mobile Host}%
\acro{MIHF}{Media Independent Handover Function}%
\acro{MLPP}{Multi Level Priority Preemption}%
\acro{Mode-S}{Mode-Select}%
\acro{MPE}{Multi Protocol Encapsulation}%
\acro{MPEG2-TS}{MPEG2 - Transport Stream}%
\acro{MPLS}{Multi Protocol Label Switching}%
\acro{NCC}{Network Control Centre}%
\acro{NET}{Network Connectivity}
\acro{NEWSKY}{NEtWorking the SKY for aeronautical communications}%
\acro{NOC}{Non-Operational Communication}%
\acro{NSIS}{Next Steps in Signalling}%
\acro{OC}{Operational Communication}%
\acro{ORP}{Oceanic, Remote and Polar}%
\acro{OSI}{Open Systems Interconnection}%
\acro{OSPF}{Open Shortes Path First}%
\acro{PDF}{Probability Density Function}%
\acro{PCE}{Predictive Capacity Estimation}%
\acro{PCN}{Pre-Congestion Notification}%
\acro{PDU}{Protocol Data Unit}%
\acro{PEP}{Performance Enhancing Proxy}%
\acro{PER}{Packet Error Rate}%
\acro{PID}{Packet Identifier}%
\acro{PLFRAME}{Physical Layer Frame}%
\acro{PLR}{Packet Loss Rates}%
\acro{QoS}{Quality of Service}%
\acro{QoS-PRN}{QoS Private Relay Nodes}%
\acro{QoS-RN}{QoS-Relay Nodes}%
\acro{RA}{Random Access}%
\acro{RBDC}{Rate Based Dynamic Capacity}%
\acro{RCB}{Random Coding Bound}%
\acro{RCS}{Return Channel via Satellite}%
\acro{RCS+M}{Return Channel via Satellite plus Mobility}%
\acro{RF}{Radio Frequency}%
\acro{RFC}{Request For Comments}%
\acro{RL}{Return Link}%
\acro{RM}{Resource Management}%
\acro{RN}{Relay Nodes}%
\acro{RP}{replicas' pointer}
\acro{RRM}{Radio Resource Management}%
\acro{RS}{Reed Solomon}%
\acro{RSVP}{Resource Reservation Protocol}%
\acro{RTT}{Round Trip Time}%
\acro{SA}{Slotted ALOHA}%
\acro{SANDRA}{Seamless Aeronautical networking Through Integration of Data Links, Radios and Antennas}%
\acro{SB}{Shannon Bound}%
\acro{S2}{Second Generation}%
\acro{SESAR}{Single European Sky ATM Research}%
\acro{SF}{Superframe}%
\acro{SIC}{Successive Interference Cancellation}%
\acro{SLS}{Service Level Specification}%
\acro{SNDCF}{Subnetwork Dependent Convergence Function}%
\acro{SNIR}{Signal to Noise and Interference ratio}%
\acro{SNR}{Signal-to-Noise ratio}%
\acro{SWIM}{System Wide Information Management}%
\acro{SYN}{Synchronization}%
\acro{SYNC}{Synchronization}%
\acro{TCP}{Transport Control Protocol}%
\acro{TCP/IP}{Transport Control Protocol / Internet Protocol Suite}%
\acro{TDMA}{Time Division Multiple Access}%
\acro{TE}{Traffic Engineering}%
\acro{TI}{Technology Independent}%
\acro{TI-SAP}{Technology Independent - Service Access Point}%
\acro{TD}{Technology Dependent}%
\acro{TL}{Transport Layer}%
\acro{TMA}{Terminal Maneuvering Area}%
\acro{ToS}{Type of Service}%
\acro{UCA}{Unconstrained Architecture}%
\acro{UDP}{User Datagram Protocol}%
\acro{VBDC}{Volume Based Dynamic Capacity}%
\acro{VDL}{VHF Digital Mode}%
\acro{VHF}{Very High Frequency}%
\acro{VoIP}{Voice over IP}%
\acro{VPN}{Virtual Private Networks}%
\acro{WG}{Working Group}%
\acro{WiMAX}{Worldwide Interoperability for Microwave Access}%
\acro{XFECFRAME}{Forward Error Correction Frame}%
\end{acronym}

%
\title{Optimum Header Positioning in \ac{SIC} based Aloha}

\author{\IEEEauthorblockN{Federico Clazzer and Christian Kissling}
\IEEEauthorblockA{German Aerospace Centre (DLR)}
\IEEEauthorblockA{Oberpfaffenhofen, D-82234, Wessling, Germany}
\IEEEauthorblockA{Email: \{federico.clazzer, christian.kissling\}@dlr.de}
}

%


\maketitle

\begin{abstract}
Random Access MAC protocols are simple and effective when the nature of the traffic is unpredictable and sporadic. In the following paper, investigations on the new \ac{ECRA} are presented, where some new aspects of the protocol are investigated. Mathematical derivation and numerical evaluation of the symbol interference probability after \ac{SIC} are here provided. Results of the optimum header positioning which is found to be in the beginning and in the end of the packets, are exploited for the evaluation of \ac{ECRA} throughput and \ac{PER} under imperfect knowledge of packets positions. Remarkable gains in the maximum throughput are observed for \ac{ECRA} w.r.t. \ac{CRA} under this assumption.
\end{abstract}


%
\IEEEpeerreviewmaketitle


\section{Introduction}

\ac{RA} \ac{MAC} protocols in the recent past have been intensively
investigated especially coupled with multi-packet per user sending
and \ac{SIC} procedure. It was first proposed in
\cite{Choudhury1983} to send the same packet twice in different
\ac{TDMA} slots. Benefits in terms of delay and throughput were
found under very moderate channel load conditions.

More recently, the idea to send multiple packets per user of
\cite{Choudhury1983} has been exploited for retrieving collided
packets with the more effective \ac{SIC} procedure in \cite{Casini2007},
\cite{Liva2011}, \cite{Kissling2011a} and \cite{Clazzer2012}.  \ac{CRDSA}
presented in \cite{Casini2007} is the first which applies
\ac{SIC} for trying to resolve packet collisions within a frame in a
\ac{SA} like \ac{RA} scheme.

\ac{CRDSA} takes from \cite{Choudhury1983} the idea to send more
than one packet instance per user for each frame. But, on the other
hand \ac{CRDSA} is designed in a way to resolve most of packet
contentions by using \ac{SIC}. The key idea of \ac{CRDSA} is to provide
in each replica the signaling information of where the other replicas of
the corresponding user are placed in the frame. This signaling information
is stored in the so called \ac{RP} section of the packet header. Every time a
packet is recovered, this information can be exploited by the \ac{SIC}
procedure for removing the signal contribution of the other replicas
from the frame, thus possibly removing its interference contribution
to other packets. The main \ac{CRDSA} advantages lie in an improved
packet loss ratio and a much higher operational throughput.

The \ac{CRA} protocol \cite{Kissling2011a} exploits the same approach
of using \ac{SIC} as \ac{CRDSA}, but in an Aloha-like \ac{RA} scheme.
Here no slots are present in the frame and thus the replicas of the
users can be placed within the frame without constraints, except that
replicas of a user may not interfere each other. The avoidance of slots
results in significant advantages such
as relaxation in synchronization requirements among users and
possibility of varying packet length without padding overhead.
\ac{FEC} in \ac{CRA} is beneficial also when
no power unbalance among users is present, unlike in \ac{CRDSA}, because partial
interference is not only possible but also more probable than
complete interference. 				
\section{ECRA and the replicas' pointers}

\begin{figure}
\centering
\includegraphics[width=8cm]{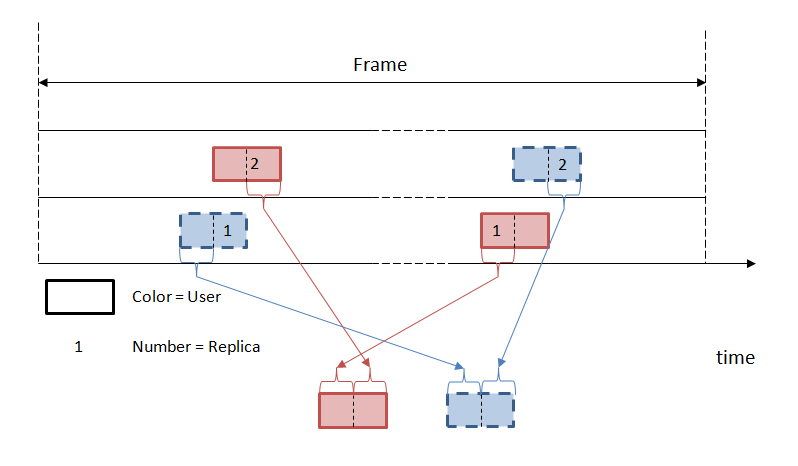}
\caption{\ac{CRA} simplest loop, and \ac{ECRA} combined packet creation}
\label{loops}
\end{figure}

In \cite{Clazzer2012} \ac{ECRA} was first presented. The protocol is an evolution of \ac{CRA}, where it is tempted to resolve \textit{loops} among packets. Both \ac{CRDSA} and \ac{CRA} are not able to resolve collisions where the packets involved belong to the same users and the inter-packet interference is too high. These situations are called \textit{loops}. Suppose we are sending two packets per user and we are using the \ac{CRDSA} protocol. Suppose further, a collision involving both the packets belonging to the two users happens. In this case there is no possibility to recover the two packets sent, and this results in a complete loss of the information stored in the users' packets. If the \ac{CRA} protocol is employed, the situation is more complex. In fact, the collision between the two users' packet can be partial, that is only a portion of the packets can be involved in the collision. For an example of loops see figure \ref{loops}. We can suppose in our case, although \ac{FEC} is used for counteracting the interference coming from the packets' collisions, the level of interference is too high for correct decoding of the packets. Also in this second case the loop cannot be resolved and the users' packets cannot be retrieved.

In \ac{ECRA} it can be observed that the partial collision of an Aloha like \ac{RA} scheme involves different parts of the packet, unlike \ac{CRDSA} where the collisions are always involving the packets entirely. Therefore, starting from the SIC algorithm of the \ac{CRA} protocol, a further step is introduced where a combined packet is created from the two (or more) user's replicas. The combined packet is composed in a way such that the resulting \ac{SNIR} is maximized, as presented in \cite{Clazzer2012}. In particular, for each symbol of one packet, the one belonging to the replica with less interference is selected to be used in the combined packet. In the case of figure \ref{loops} we can see how the combined packet of \ac{ECRA} will be created. Due to the higher \ac{SNIR} further packets can be retrived from the frame, increasing the throughput performance and decreasing the \ac{PER} w.r.t. \ac{CRA}. An increase up to 26\% in the peak throughput is achieved by \ac{ECRA} w.r.t. \ac{CRA} as shown in \cite{Clazzer2012}. The key assumption there is the perfect knowledge of all the packets positions within the frame after the first step of \ac{SIC}. Differently from \cite{ZigZag}, where an iterative \textit{chunk-by-chunk} decoding between the collided packets is employed, in \ac{ECRA} an entire combined packet is constructed and the decoding attempted on it in one step. Moreover, in \ac{ECRA} the replicas generation is made regardless of the decoding success, while in \cite{ZigZag} only the collided packets are replicated.

The aim of this work is two fold. First to investigate what is the best header and thus \ac{RP} positioning, and second to evaluate the behavior of \ac{ECRA} in terms of throughput when the aforementioned assumption is removed and the best \ac{RP} positioning is employed. 					
\section{Optimum replicas' pointers position}

In \ac{RA} protocols without slots and with \ac{SIC} such as \cite{Kissling2011a} and \cite{Clazzer2012}, the packets are sent randomly within a frame. Therefore, when a collision between two packets happens, the quantity of symbols involved in that collision is uniformly distributed between 1 and the length of the packet $P_{len}$. If we consider the whole set of collided packets, it can be shown that the probability of collision for every symbol within the generic packet, is equal for all the symbols, $p_i=p,$ for $i=1,...,P_{len}$. Since every symbol has the same probability to face interference, there is no way to optimize the header positioning in the packet in order to minimize the probability to have interference in these sensitive sections.

However an interesting question is, whether sections of the packets can be found which are less subject to interference, if interference cancellation is applied. To address this question let us first consider the simplest interference case of only two packets interfering each other.

\subsection{Case I: Two interfering packets}

First we assume that the colliding packets are symbol synchronized, hence, when a symbol is involved in a collision it is entirely collided. Second we assume perfect \ac{SIC}, therefore all the decodable packets are correctly received and ideally removed from the frame. Third, the received packet power is the same for all the packets. Fourth, the decoding threshold is assumed to be the Shannon capacity limit $SNIR_{SHA}$ and thus, we are considering Gaussian inputs. According to \cite{Kissling2011a} and \cite{Clazzer2012} we can write the \ac{SNIR} decoding threshold as $SNIR_{SHA} = 2^R-1$, given the rate $R$. We can also write the \ac{SNIR} of user $u$ and replica $r$ as $SNIR_{u,r}=\frac{P}{x_{u,r}\cdot P + N}$. The quantity $x_{u,r}$ is the interference ratio suffered by the packet of user $u$ and replica $r$, e.g. $x_{u,r}=1$ means that one packet is entirely colliding with the given packet. Every packet with $SNIR_{u,r}\geq SNIR_{SHA}$ is correctly received, thanks to the fourth hypothesis. We are interested in determining the quantity $x^*$ which is the maximum percentage of interference for which a packet is still decodable at the receiver. This value can be derived by the equation

\begin{equation}
\label{21}
SNIR_{SHA}=SNIR_{min} \Rightarrow 2^R-1=\frac{P/N}{x^*\cdot P/N + 1},
\end{equation}

where we can define \ac{SNR} as $SNR=P/N$. This leads to

\[
x^*=\frac{1}{2^R-1}-\frac{1}{SNR}.
\]

Thus, $x^*$ can be determined when the rate $R$ and the packet's $SNR$ are defined. For a better understanding of the scenario, let us divide the two interfering packets case in two subcases $x^*\geq0.5$ and $x^*<0.5$.

\subsubsection{\underline{$x^*\geq0.5$}}

We are interested in determining the number of possible cases $N_{int}$, where two packets are interfering. Since the packets are symbol synchronized, this number is finite and is $N_{int} = 2 \cdot P_{len} - 1$. This value can be derived observing that at least one symbol of the two packets needs to be interfered. In order to ensure this, we have $2 \cdot (P_{len} - 1) + 1$ possible interference cases among two packets, which lead to the value of $N_{int}$.

Let us now define $L=\lfloor P_{len}\cdot x^* \rfloor + 1$, where $L$ represents the minimum number of symbols that make a packet undecodable. Due to the second hypothesis, all the packets with $0,...,L-1$ symbols interfered were already correctly decoded. Therefore, after the \ac{SIC} process the number of possible cases $N_{int_{after-SIC}}$, where two packets are interfering reduces to
\[
N_{int_{after-SIC}}=2 \cdot (P_{len} - L) + 1.
\]

Intuitively, the formula derives directly from the previous reasoning, where instead of one, at least L symbols need to be interfered.

We now want to determine the probability of interference for each symbol in the packet $p_i,$ for $i=1,...,P_{len}$. Since $x^*\geq0.5$, it follows that $L>P_{len}/2$ and therefore some of the central symbols of the packet must face interference with probability equal to 1. The exact position of these symbols within the packet is related to $L$. We can observe that the least interference possible is when exactly $L$ symbols are involved in a collision. In this case the first $P_{len} - L$ symbols of the first packet collided and the last $P_{len} - L$ symbols of the second packet collided. Call $S_{c_1}$ the set of collided symbols of the first packet, and $S_{c_2}$ the set of collided symbols of the second packet. Explicitly

\[
S_{c_1} = \left\{ 1,...,L \right\} \ and \ S_{c_2} = \left\{ P_{len}-L+1,...,P_{len} \right\}.
\]

The intersection of symbols involved in the collision from both packets, are the ones with probability equal to 1 to be interfered. This can be derived by the observation that in all the other possible collisions, more than $L$ symbols are involved in the interference. Call $S_c$ the set given by $S_{c_1} \cap S_{c_2}$,

\[
S_c=S_{c_1} \cap S_{c_2}=\left\{ P_{len}-L+1,...,L \right\}.
\]

For every symbol outside $S_c$, the probability of interference is $<1$. The symbol immediately before symbol $P_{len}-L+1$, called $s_{P_{len}-L}$, is involved in collisions in all the cases except one. The case where this symbol is not involved in the collision is when exactly the last $L$ symbols of the packet are interfering. In a similar way, the symbol immediately after symbol $L$, called $s_{L+1}$, is also involved in all the cases except one. If we iterate this reasoning until symbol $s_1$ in the backward direction and symbol $s_{P_{len}}$ in the forward direction, we find $p_1 = p_{P_{len}} = \frac{P_{len}-L+1}{2 \cdot (P_{len} - L) + 1}$.

We are now able to formulate the interference probability $p_i$ for each symbol in the packet, $i=1,..,P_{len}$ as

\begin{equation}
\label{31}
p_i = \left\{
\begin{array}{rl}
\frac{P_{len}-L+i}{2 \cdot (P_{len} - L) + 1} & \mbox{for $i=1,...,P_{len}-L$}\\
1                & \mbox{for $i=P_{len}-L+1,...,L$}\\
\frac{2 \cdot (P_{len} - L) + 1 - (i-L)}{2 \cdot (P_{len} - L) + 1} & \mbox{for $i=L+1,...,P_{len}$}
\end{array}
\right .
\end{equation}

\subsubsection{\underline{$x^*<0.5$}}

If we now consider the case of $x^*<0.5$, the first observation is that the total number of cases where two packets are interfering $N_{int}$ is again, $N_{int_{after-SIC}}=2 \cdot (P_{len} - L) + 1$. It should be noted that the expression of $N_{int}$ is the same, but the actual value has changed because $L$ is different from the previous case. The second observation is that $S_c=\varnothing$, because

\begin{equation}
\label{32}
L<P_{len}-L+1, \ for \ each \ L \ with \ x^*<0.5
\end{equation}

The demonstration is the following, $L=\lfloor P_{len}\cdot x^* \rfloor + 1$, leads to $L\leq P_{len}/2$. Call the maximum of $L$, $L_{max}$

\begin{equation}
\label{33}
L_{max}=\frac{P_{len}}{2}.
\end{equation}

For $L=L_{max}$ we are maximizing the left hand and minimizing the right hand of inequality \eqref{32},

\[
L_{max}<P_{len}-L_{max}+1 \Rightarrow_{with \eqref{33}} \frac{P_{len}}{2}<\frac{P_{len}}{2}+1.
\]

Therefore, in this second case, no symbols have probability 1 to face interference. The maximum number of possible cases where a symbol is involved in collisions is $P_{len}$, and the symbols that are involved in all these cases are $s_i$ with $i=L,...,P_{len}-L+1$. This result can be found following the same reasoning presented in the previous paragraph with the appropriate differences. Moreover, the first and last symbol interference probability that can be computed applying the same procedure, is again $p_1 = p_{P_{len}} = \frac{P_{len}-L+1}{2 \cdot (P_{len} - L) + 1}$.

We are now able to formulate the interference probability $p_i$ for each symbol in the packet, $i=1,..,P_{len}$ as

\begin{equation}
\label{34}
p_i = \left\{
\begin{array}{rl}
\frac{P_{len}-L+i}{2 \cdot (P_{len} - L) + 1} & \mbox{for $i=1,...,L-1$}\\
\frac{P_{len}}{2 \cdot (P_{len} - L) + 1}     & \mbox{for $i=L,...,P_{len}-L+1$}\\
\frac{2 \cdot (P_{len} - L) + 1 - (i-L)}{2 \cdot (P_{len} - L) + 1} & \mbox{for $i=P_{len}-L+2,...,P_{len}$}
\end{array}
\right .
\end{equation}

\begin{figure}
\centering
\includegraphics[width=8cm]{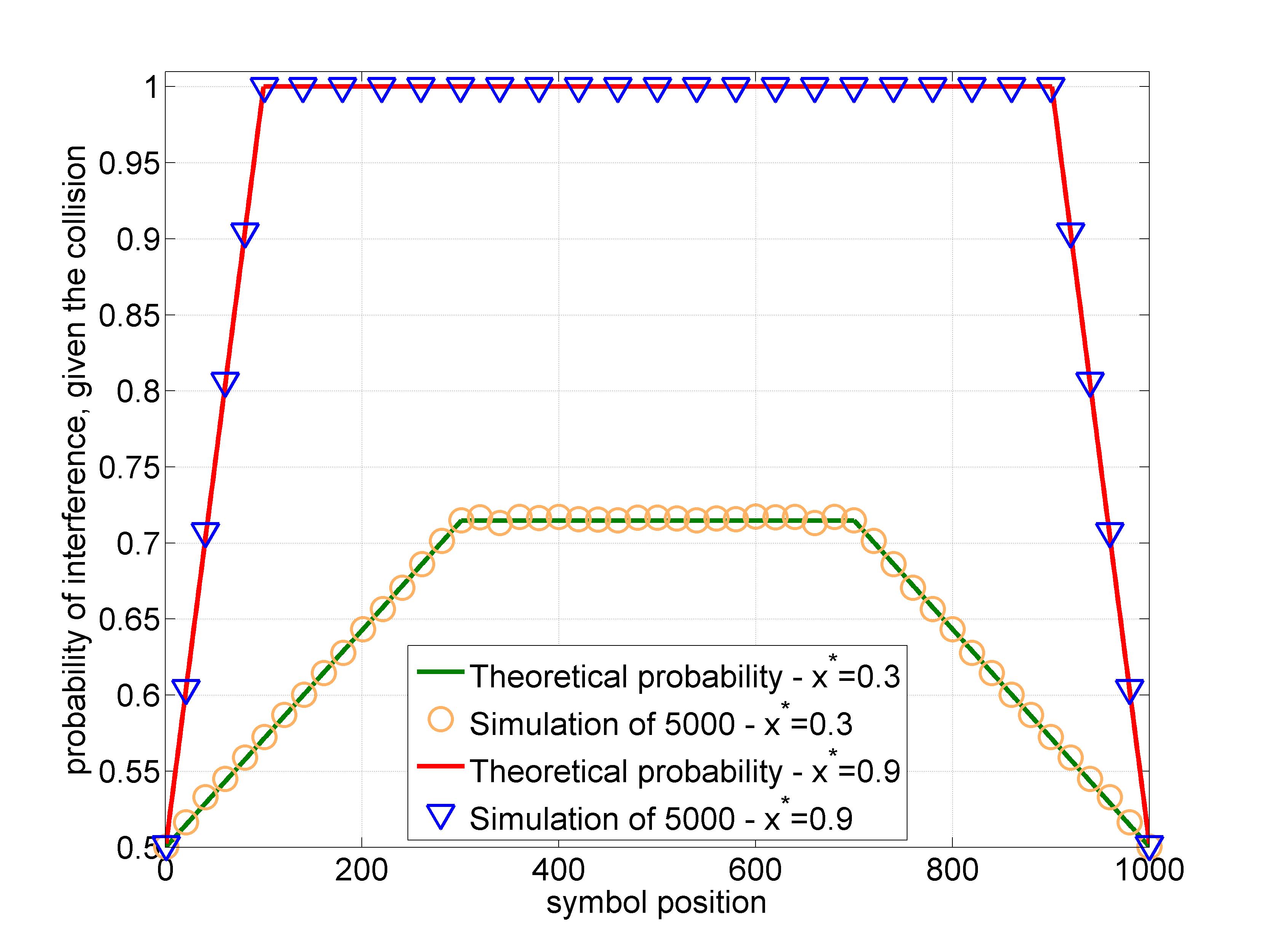}
\caption{Collision of 2 packets after SIC, theoretical probability and empirical probability comparison}
\label{2Pack}
\end{figure}

In figure \ref{2Pack} the comparison between the average empirical probability of symbol interference among 5000 collisions and the theoretical probability is presented. A packet length $P_{len}=1000$ symbols and two representative thresholds, $x^*=0.9$ and $x^*=0.3$, have been selected. In order to have a quantitative evaluation of the difference between the empirical and the theoretical probabilities, the \ac{KL-d} has been evaluated. In particular

\[
KL_d = \sum_t V(t) \ln \frac{V(t)}{Q(t)},
\]

where $t=1,...,P_{len}$, $V$ is the theoretical probability and $Q$ is the empirical probability. The useful property of the \ac{KL-d} is that $KL_d = 0$ if and only if $V(t)=Q(t)$ for $\forall t$. Therefore, the closer $KL_d$ is to 0, the more the empirical and theoretical probabilities are matching. In our case the $KL_d$ distance gets $KL_d = -0.7019$ for the case of $x^*=0.3$ and $KL_d = -0.3053$ for $x^*=0.9$. In both cases \ac{KL-d} is very close to 0, which underlines the validity of the simulation results compared to the mathematical analysis.

In the case of $x^*=0.9$, the symbols at the very beginning and the symbols at the very end of the packet have quite half of the probability to face interference with respect to the central symbols. In the case of $x^*=0.3$ this ratio reduces to $0.30$, but it is still remarkable.

\subsection{Case II: Three or more interfering packets}

\begin{figure}
\centering
\includegraphics[width=8cm]{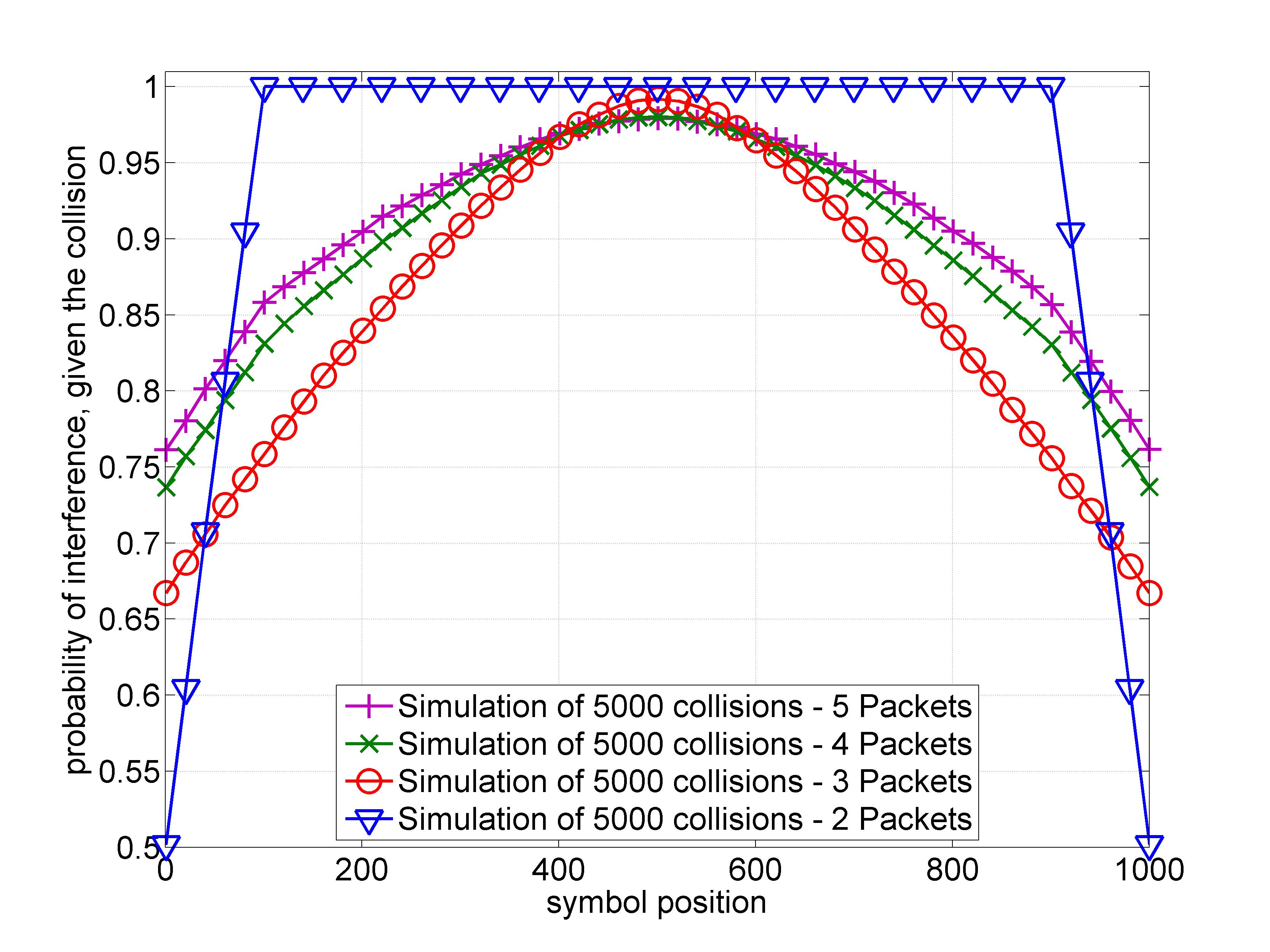}
\caption{Collision of 2, 3, 4 and 5 packets after SIC for $x^*=0.9$, $P_{len}=1000$ symbols and $L=901$ symbols - simulation results}
\label{2-3-4-5Pack_0_9}
\end{figure}

The evaluation of the cases of three or more packets interfering is carried out with the help of simulations. Moreover, some qualitative remarks are provided in the following. If we start considering the case of $x^*\geq 0.5$ and interference of three or more packets, we can make two considerations: first, there can be cases where the symbols in the packet center are free of interference, second, symbols in the beginning and in the end will have a higher probability of interference w.r.t. the two packets interference case as is shown in the simulations presented in figure \ref{2-3-4-5Pack_0_9}. These two remarks are the effects of the same cause, if the collision involves more than two packets, in some cases the central section of one or more packets can be free from interference, but at the same time higher interference must be found at the beginning or at the end of the packet. The case of $x^*< 0.5$ is slightly different, in fact not only the beginning and ending symbols will have an increased probability of interference w.r.t. the two interfering packets case, but also the central symbols will face interference more frequently. This is due to the increased number of combinations of interference that can be expected in the case of more than two packets interfering which lead to increased probability of interference among all the symbols in the packet.

\begin{figure}
\centering
\includegraphics[width=8cm]{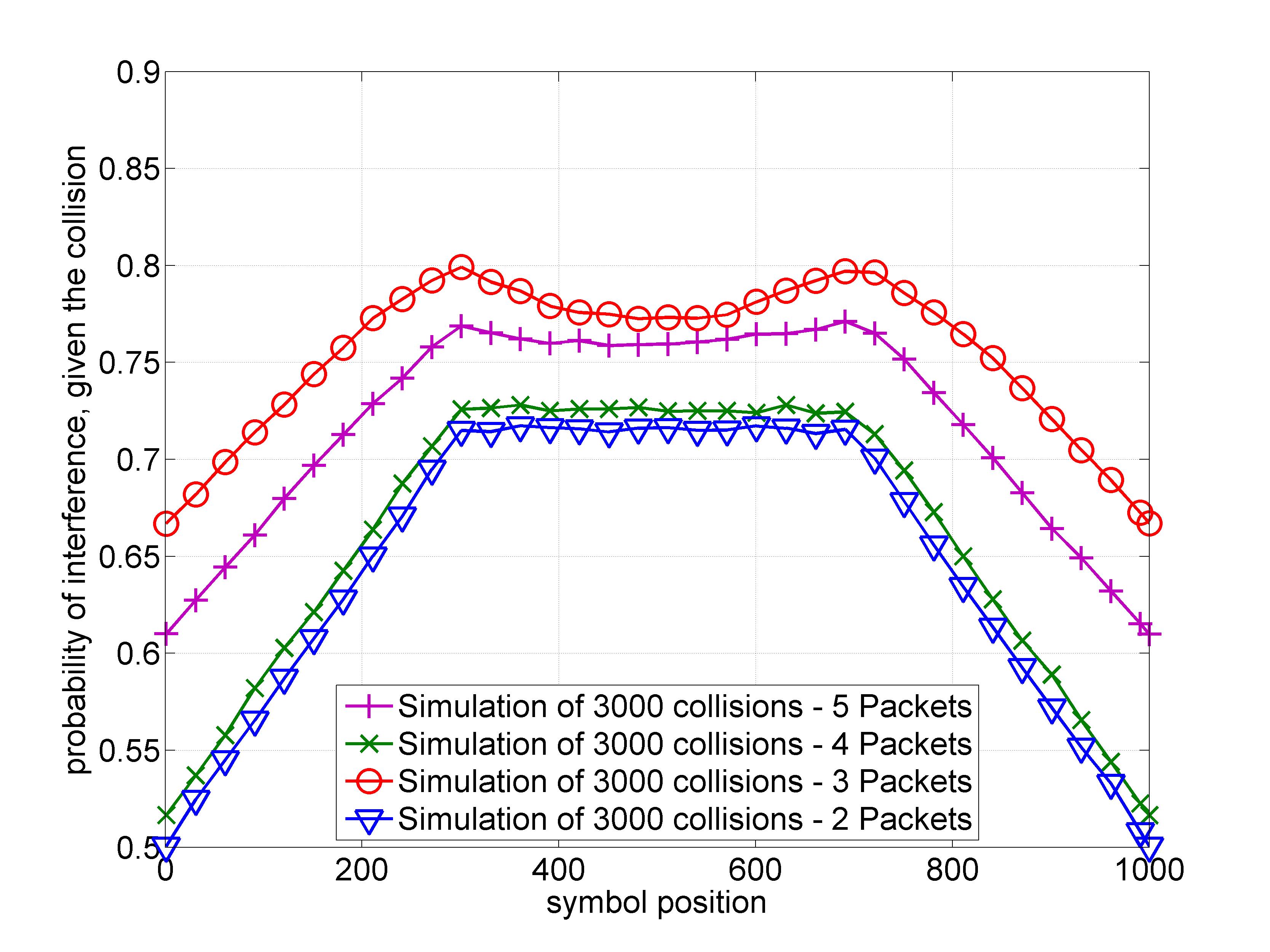}
\caption{Collision of 2, 3, 4 and 5 packets after SIC for $x^*=0.3$, $P_{len}=1000$ symbols and $L=301$ symbols  - simulation results}
\label{2-3-4-5Pack_0_3}
\end{figure}

The symbol probability of every packet symbol for 2, 3, 4 and 5 packets colliding with $x^*=0.9$ is presented in figure \ref{2-3-4-5Pack_0_9} while the same symbol probability but with $x^*=0.3$ is presented in figure \ref{2-3-4-5Pack_0_3}. All the aforementioned qualitative remarks are confirmed by the simulations. In figure \ref{2-3-4-5Pack_0_9} it can noted that already for three packets colliding, the probability of interference for every symbol is lower than 1. Furthermore, the beginning and ending symbols have more probability to be interfered w.r.t. the two packets colliding case. In case of $x^*=0.3$, the results presented in figure \ref{2-3-4-5Pack_0_3} show an increase of symbol interference probability for all the symbol positions. It can be also noted in the case of $x^*=0.3$, that for 3 packets colliding two peaks of interference probability can be found around $L$ and $P_{len}-L+1$. This effect is smothered in case of 5 packets colliding. This fact can be qualitatively explained in the following way: when 3 packets collide and the minimum level of interference is considered, in one packet only one of the two border parts, the beginning section or the ending section, can be involved in the collision. Otherwise, when 4 packets collide the constraint disappears and both sections can be interfered.

A fundamental conclusion can be drawn from the results presented here, regardless the number of packets involved in a collision and regardless the decoding threshold, the symbols in the beginning and in the end of a packet are less frequently involved in collisions. The property can be exploited for a smart header positioning within the packet. 							
\section{Numerical Results}

Thanks to the result of the previous section, we know that positioning the header and therefore the \ac{RP} in the beginning or in the end of a packet is the best choice for minimizing the probability of interference on it after the \ac{SIC} process.

In the following results are provided for the probability of interference in the packet headers after the \ac{SIC} procedure of \ac{CRA}. The aim is first to show how the symbol interference probability evaluated in the previous section translates to packet header interference probability after \ac{SIC} and second, to use this statistics for the \ac{ECRA} throughput and \ac{PER} simulations.

\subsection{Header interference probability}

The header length $H_l$ [symbols] is fixed to a certain value, in this example 10\% of the packet length. Bigger headers will have too much impact on the overhead of the protocol, hence this choice can be seen as a worst case assumption on the header length. The header has a \ac{CRC} code through which the decoder is able to recognize any impairment on this section of the packet. We assume that if one or more symbols of the packet header face interference, the header is not decodable and therefore we cannot retrieve the information about the replicas positioning of the user. The simulation assumptions are the following, the rate $R=2$ is selected and the \ac{SNR} is $SNR=10$ dB equal for each user generating traffic. The frame duration $T_f$ is $T_f=100$ ms and the symbol duration $T_s$ is $T_s=1$ $\mu$s. The packet length is $L_p=1000$ bits and is equal for each user. If we fix the number of users $N_u$, the offered traffic load $G$ is computed as $G=\frac{N_u\cdot L_p\cdot T_s}{T_f\cdot R}$.

Given $R$ and $L_p$ we can calculate the symbol length of the packet as $P_{len}=L_p/R=500$ symbols. The number of replicas sent by each user within a frame is $d=2$. The maximum number of \ac{SIC} iterations $I_{max}$ is chosen to be $I_{max}=10$. In this way, a first step of \ac{SIC} procedure is performed on the user packets at the decoder. The packets which have a sufficient \ac{SNIR} can be correctly decoded and removed from the frame. In our simulations the Shannon capacity limit is assumed as decoding threshold according to \eqref{21}. After the \ac{SIC} procedure the packets which are still not decoded are checked and the ratio between the number of remaining packets and the number of packets with interference in the header is evaluated. This ratio is the empirical probability of having interference in the packet header after the \ac{SIC} procedure.

In figure \ref{head_prob} the results for different header positions and different header lengths are provided. Two header lengths are assumed $H_{l-1}=10\%\cdot P_{len}$ (continuous curves) and $H_{l-2}=5\%\cdot P_{len}$ (dotted curves). The scenarios with $H_{l-1}$ can be seen as a worst case, in fact having a header which occupies more than 10\% of a packet is quite inefficient and would be an overhead hardly acceptable, while the choice $H_{l-2}$ appears more appealing. As expected the worst case is when the header is placed in the central part of the packet. In this case we have the highest probability that the header symbols face interference, (see figure \ref{2-3-4-5Pack_0_3}). If we call $p_{h-int}$ the probability of interference in the header, in this first case $p_{h-int} > 0.7$ for all the values of offered traffic load $G$ for $H_{l-1}$. For low values of $G$, if we decrease the header length to $H_{l-2}$ an observable difference in $p_{h-int}$ is found which decreases as $G$ grows.

When we move the header to the beginning of the packet, the interference probability decreases by more than 10\% for both header lengths cases and for $0.1\leq G \leq 0.6$ w.r.t. to the cases of a header in the center. Also for this second case, the decrease of the header length is beneficial from the probability of header interference point of view. Thanks to the previous section we are ensured that no other positioning of the header can achieve better probability of interference $p_{h-int}$. Moreover, thanks to the symmetry of the symbol interference probability w.r.t. the center of the packet, we know that positioning the header at the end of the packet will achieve the same probability of interference $p_{h-int}$.

Since the information carried by the header on the user replicas position is fundamental for \ac{ECRA}, it can be thought that replicating this information could improve the reliability. Therefore, two more cases are shown in the figure. The first supposes that the header is placed twice per packet in the beginning and in the end. Here the total overhead due to the header replication is 20\% of $P_{len}$ assuming $H_{l-1}$ and 10\% of $P_{len}$ assuming $H_{l-2}$. The second additional case assumes that the header is replicated three times inside each packet: at the beginning, in the center and at the end. In this last case the total header overhead is 30\% of $P_{len}$ assuming $H_{l-1}$ and 15\% of $P_{len}$ assuming $H_{l-2}$. The advantage is clear, the header cannot be correctly decoded only if at least one symbol for each of the two or three places where the header is replicated faces interference. Conversely to the other two cases, only one or more symbols interfering in one of the header locations will not prevent the retrieval of the header information and the \ac{RP}. On the other hand, given a fixed packet length, the replication of the header increases the overhead and decreases the throughput. The simulations of the next section will help to understand the benefits of the header replication. If we consider both cases for $H_{l-1}$, the header interference probability is below 0.2 for $G\leq 0.3$ and remains below 0.5 until $G=0.55$. The difference between the two cases is observable for $G\geq0.4$ but very limited in absolute terms ($<0.4\%$). Moreover, if we consider both the cases for $H_{l-2}$, the header interference probability is lower than 0.2 for $G\leq 0.4$ and remains under 0.5 until $G=0.65$. These results are very promising, especially for the case of header positioning in the beginning and in the end of the packet. The very limited decrease of header interference probability of the three headers case w.r.t the two headers case, cannot counteract the decrease of information sent per packet due to the increase of symbols required by the replication of the header, as will be shown in the next subsection.

\subsection{ECRA throughput simulations without perfect packets position knowledge}

\begin{figure}
\centering
\includegraphics[width=9.5cm]{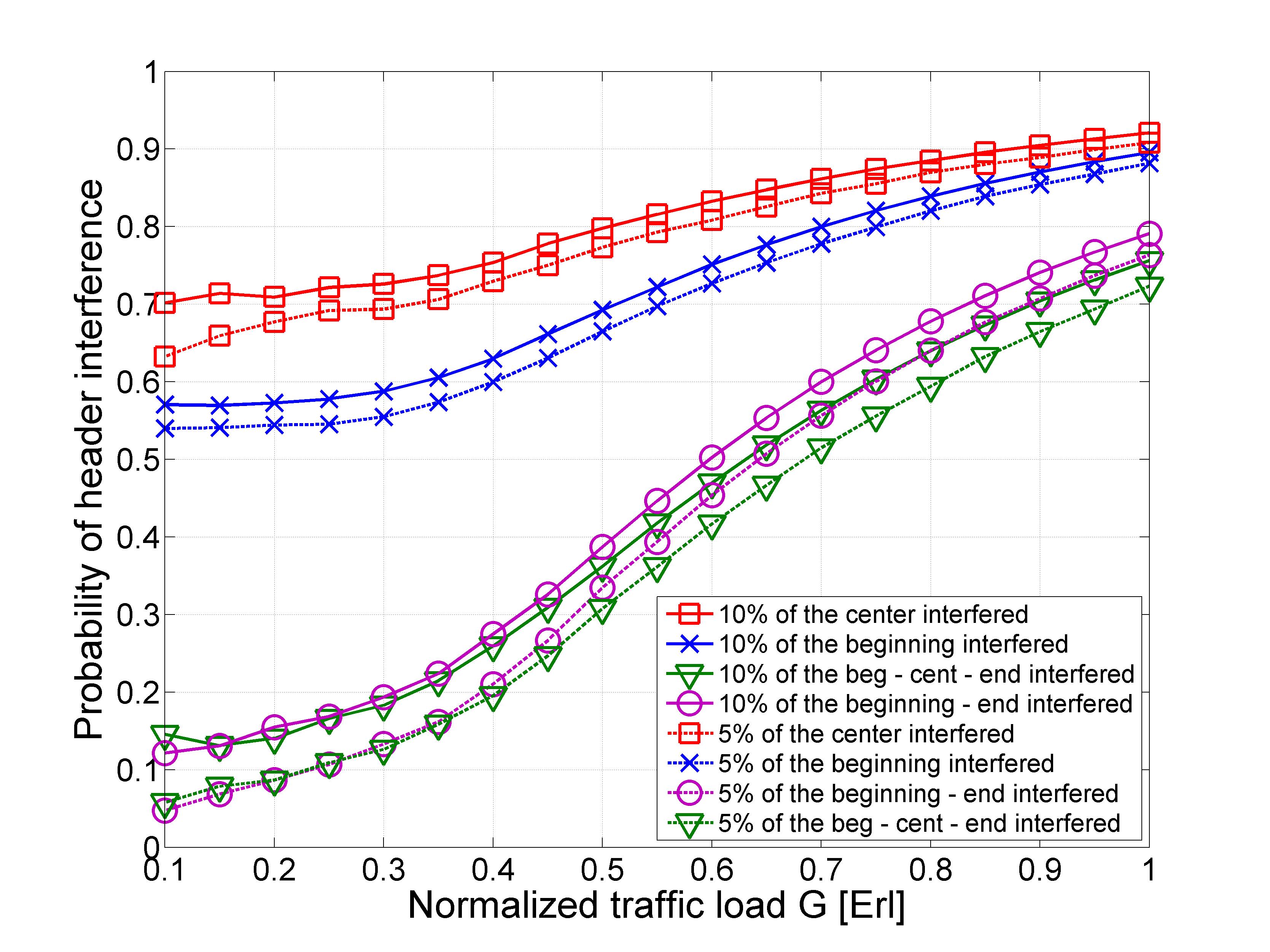}
\caption{Header interference probability for different header positions and different header lengths}
\label{head_prob}
\end{figure}

After the first step where \ac{SIC} is performed as in \ac{CRA}, the \ac{ECRA} protocol tries to create a mix packet from the two user replicas which has the lowest level of interference possible and performs the decoding on it. If it is successful, a second round of \ac{SIC} is done. In \cite{Clazzer2012}, it was assumed that although the remaining packets were not decoded, it was always possible to retrieve the information of their position. Indeed this is not always the case, and here this assumption is removed. Instead, we assume that the information in the \ac{RP} can be recovered with probability $p_{h-int}$ (simulated in figure \ref{head_prob}) because $p_{h-int}$ represents the probability of no interference in the header(s) where the \ac{RP} is stored. Therefore with probability $p_{h-int}$ we are able to create the mix packet and to try the decoding. Since $p_{h-int}$ depends strongly on the simulation parameters, the throughput and \ac{PER} simulations of \ac{ECRA} will use the same parameters listed in the previous subsection.

The average packet error rate $\overline{PER}$, is evaluated as:

\[
\overline{PER} = \frac{P_{err}}{N_u \cdot N_f}
\]

where $P_{err}$ is the number of lost packets at the receiver side, and $N_f$ is the number of simulated frames for the corresponding
$G$ which is $N_f=10^3$. The average throughput $\overline{T}$ is defined as the probability of successful reception of a packet, multiplied by the offered traffic load $G$. The average throughput here is related to the logical throughput, i.e. user packets, whereas the physical throughput would also consider the number of replicas generated per packet. Since the $\overline{PER}$ represents the average probability of a packet error, $\overline{T}$ is computed in the following way:

\begin{equation}
\label{41}
\overline{T} = \left[(1-\overline{PER}) \cdot G \right] \cdot \left[1-(n-1)\cdot \frac{H_l}{P_{len}} \right].
\end{equation}

Where $n$ is the number of headers per packet. With equation \eqref{41} we are taking into consideration the loss of information sent per packet due to the replication of the header. It can be noted that for $\lim_{H_l\rightarrow0} \overline{T} = (1-\overline{PER}) \cdot G$, e.g. the smaller the packet header w.r.t. to the packet length, the lower is the impact in terms of throughput of the header replication.

\begin{figure}
\centering
\includegraphics[width=9.5cm]{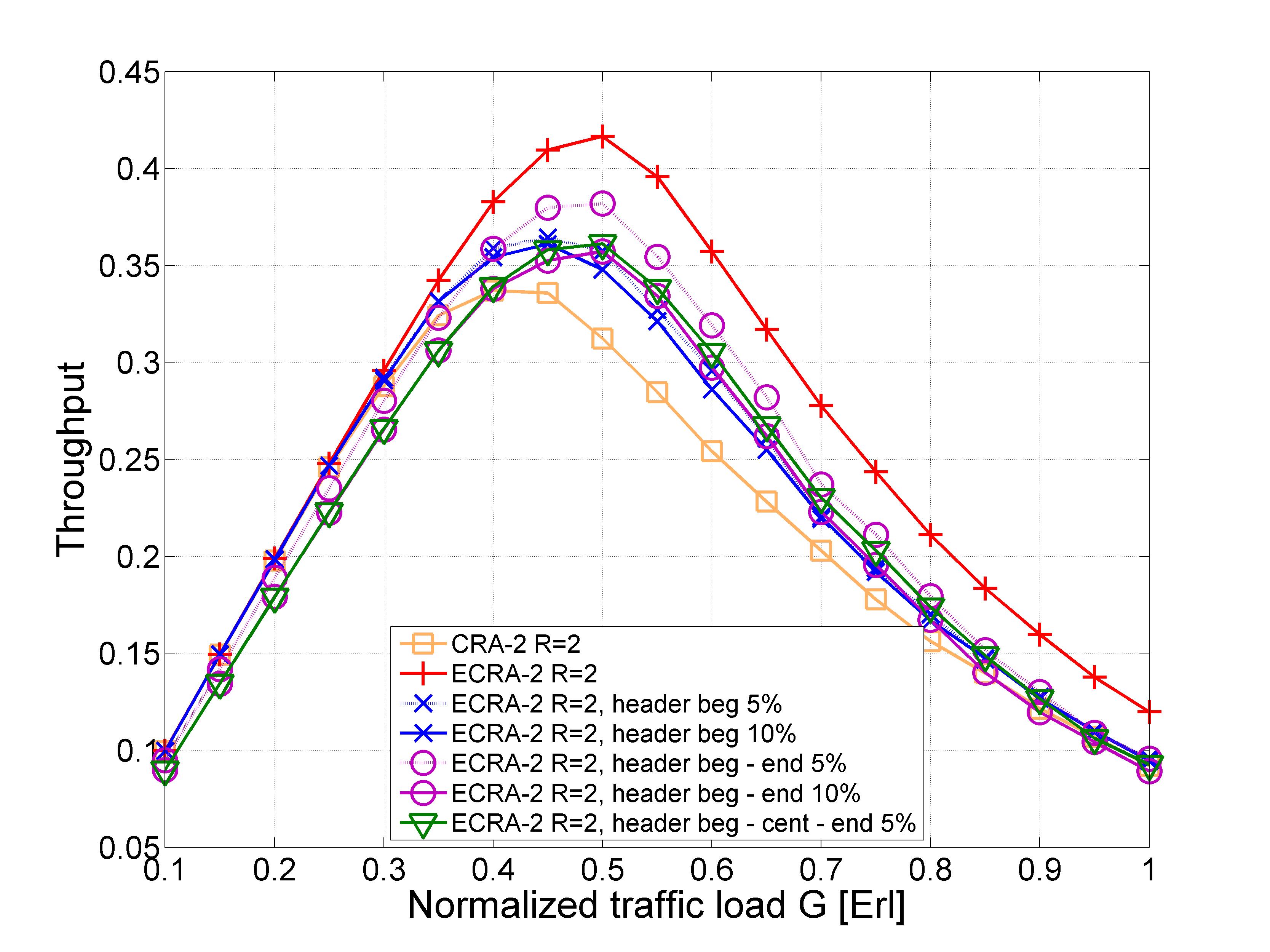}
\caption{ECRA-2 Throughput simulations}
\label{Ecra_t}
\end{figure}

In figure \ref{Ecra_t} the throughput simulations for different header positions and header lengths are provided. In terms of maximum throughput $T_{max}$, the best choice appears positioning the header twice per packet at the beginning and at the end. In this case with the smaller header length $H_{l-1}$ the increase of $T_{max}$ w.r.t. \ac{CRA} is 13\%, while in the case of perfect knowledge of the packet positions the increase is 23\%. For all the other header positions and header lengths, the gain of \ac{ECRA} is reduced to 7\%. Please note that when the header is replicated within the packet, the throughput performance for $G<G_{max}$ are always slightly worse than the cases with no header replication. This is due to the decrease of information sent per packet. 				
\section{Conclusions}

In this work, the probability of symbol interference after \ac{SIC} procedure was mathematically derived for the case of two colliding packets and simulated up to 5 packets colliding. Interestingly, it is shown that regardless of the number of packets involved in the collision, the beginning and the end part of the packet has the lowest probability to face interference. Exploiting this property, the packet header and the \ac{RP} can be placed in the beginning of the packets. Moreover, it is shown that replicating twice the header in the beginning and in the end of each packet is beneficial also from the \ac{ECRA} throughput point of view although less information per packet is sent. \ac{ECRA} throughput simulations are provided when the hypothesis of perfect knowledge of the packets position within the frame is removed. \ac{ECRA} in the case of header positioning in the beginning and in the end of the packets, showed a throughput gain of 13\% w.r.t. \ac{CRA} if the header has a length of 5\% of the packets length. Moreover, it is shown that if the length of the header can be further decreased, a gain up to 23\% in the maximum throughput can be reached by \ac{ECRA} w.r.t. \ac{CRA}.  								

\ifCLASSOPTIONcaptionsoff
  \newpage
\fi

\bibliographystyle{IEEEtran}
\bibliography{IEEEabrv,References}
%





\end{document}